\documentclass[prl,twocolumn]{revtex4}
\usepackage{graphicx}
\begin{document}

\newcommand {\beq} {\begin{equation}}
\newcommand {\eeq} {\end{equation}}
\newcommand {\bqa} {\begin{eqnarray}}
\newcommand {\eqa} {\end{eqnarray}}
\newcommand {\xx} {\ensuremath{{\bf{x}}}}
\newcommand {\kk} {\ensuremath{{\bf{k}}}}
\newcommand {\kp} {\ensuremath{{\bf{k'}}}}
\newcommand {\ca} {\ensuremath{c^{\dagger}}}
\newcommand {\no} {\nonumber}
\newcommand {\sib} {\ensuremath{\overline{\psi}}}
\newcommand {\br} {\ensuremath{\begin{array}{clrr}}}
\newcommand {\er} {\ensuremath{\end{array}}}
\newcommand {\MM} {\ensuremath{{\bf{M}}}}
\newcommand {\A} {\ensuremath{{\bf{A}}}}
\newcommand {\DD} {\ensuremath{{\bf{D}}}}
\newcommand {\LL} {\ensuremath{{\bf{L}}}}
\newcommand {\UU} {\ensuremath{{\bf{U}}}}
\newcommand {\WW} {\ensuremath{{\bf{W}}}}
\newcommand {\VV} {\ensuremath{{\bf{V}}}}
\newcommand {\dn} {\ensuremath{\downarrow}}
\newcommand {\gr} {\ensuremath{{\bf {G}}}}
\newcommand {\up} {\ensuremath{\uparrow}}
\newcommand {\Ds} {\ensuremath{\Delta^*}}
\newcommand {\dd} {\ensuremath{\Delta}}
\newcommand {\qq} {\ensuremath{{\bf{q}}}}
\newcommand {\QQ} {\ensuremath{{\bf{Q}}}}
\newcommand {\KK} {\ensuremath{{\bf{K}}}}
\newcommand {\dl} {\ensuremath{\Delta_0}}
\newcommand {\ep} {\ensuremath{\epsilon}}
\newcommand {\Do} {\ensuremath{\Delta_0}}
\newcommand {\ms} {\medskip}

\title{BCS-BEC crossover with unequal mass fermions}
\author{Roberto B. Diener}
\author{Mohit Randeria}
\affiliation{Department of Physics, The Ohio State University, Columbus, OH 43210}
\date{\today}
\begin{abstract}
We investigate the crossover from BCS pairing to molecular BEC in an atomic gas
with two fermion species with masses $m_\up \ne m_\dn$ tuned through a 
Feshbach resonance. We present results for the $T=0$ equation of state as a function of the scattering length
including the effects of Gaussian fluctuations about the mean field ground state. 
We compute the ground state energy as a function of $m_\up/m_\dn$ at unitarity and find excellent agreement 
with the quantum Monte Carlo result for $m_\up/m_\dn = 6.67$ for a $^{40}$K-$^6$Li mixture.
We show that the dimer scattering length in the BEC limit as a function of $m_\up/m_\dn$ 
compares well with exact four-body results of Petrov {\it et al}. We also derive the condition for
trapping frequencies to obtain an unpolarized gas in a harmonic trap.
\end{abstract}
\maketitle
   
The crossover~\cite{Leggett,NSR,engelbrecht} of collective BCS pairing to the BEC of tightly bound molecules
is of great interest in diverse areas of physics ranging from condensed matter to nuclear and 
high-energy physics. The unitary scattering regime which lies in the middle of the crossover in three 
dimensions permits us to gain a deeper understanding of superfluidity in a strongly interacting 
Fermi system. Recent advances in atomic cooling and trapping and the use of a Feshbach resonance to tune the
interactions through unitarity have led to the experimental realization~\cite{equal mass experiments} of the BCS-BEC crossover
in ultracold atomic gases.

In this letter we turn to the question of pairing and superfluidity in a mixture of two species of fermions with
\emph{different} masses. There are several motivations for undertaking such an investigation.
First, from an experimental point of view, interspecies Feshbach resonances have already been observed in mixtures of 
fermionic $^6$Li and $^{40}$K~\cite{interspecies Feshbach}. This provides a whole new system in which strong 
interactions in the BCS-BEC crossover can be studied. 
Second, in high energy studies of pairing, such as in color superconductivity, 
the pairing naturally occurs between fermions with different masses~\cite{colorSC}.
Third, from a general many-body physics point of view it is interesting to ask
if the unequal ``spin" $\up$ and $\dn$  mass, which breaks
``time-reversal'' symmetry in condensed matter systems, affects the pairing of 
$|\kk,\up\rangle$ and $|-\kk,\dn\rangle$ states.

Here we present results for the $T=0$ equation of state for an unpolarized system
for arbitrary scattering length and mass ratio.
We generalize to the case of unequal masses a functional integral formulation~\cite{Paper1} 
that goes beyond the mean field theory results 
reported recently~\cite{SadeMelo} and incorporates the effects of quantum fluctuations. Specifically, our approach
includes the energy of the zero point motion of the collective mode (i.e., the Goldstone mode in the broken symmetry superfluid state)
as well as the effects of virtual scattering of two-particle excitations. 

Our results (Figs.~1,2,3) are significantly different from mean field theory and also from those of the 
$1/{\cal N}$ approximation~\cite{1overN}, but are in very good agreement with
the few known exact results. At unitarity, our results compare well (Fig.~2) with
quantum Monte Carlo results that are available for two values of the mass ratio: equal masses~\cite{Giorgini}
and Li-K mixtures~\cite{Carlson}. In the BEC limit we find that our dimer scattering length as a function of
the mass ratio (Fig.~3) is in very good agreement with the exact four-body results~\cite{Petrov}. 
In the BEC regime we also find approximately 90\% of the the Lee-Yang-Huang correction caused by quantum depletion of the condensate.

The Hamiltonian density for the system is given by
\beq
H=\sib_{\sigma}(x)\left[\frac{-\nabla^2}{2m_\sigma}-\mu_\sigma \right]\psi_{\sigma}(x)-g\sib_{\up}(x)\sib_{\dn}(x)\psi_{\dn}(x)\psi_{\up}(x)
\label{hamiltonian}
\eeq
where we use ``spin" $\sigma = \uparrow, \downarrow$ to label the two fermion species with masses $m_\up \ge m_\dn$. 
The short-range attraction~\cite{wide resonance} with ultraviolet cutoff $\Lambda$ has a strength $g(\Lambda)$ that is 
related to the s-wave scattering length $a_s$ via ~\cite{engelbrecht} 
$m/4\pi a_s = -1/g(\Lambda) + \sum_{|\kk| < \Lambda} m / k^2$, with $m$ defined below.
We set $\hbar = 1$ and work in a box of unit volume.

We have two chemical potentials $\mu_\sigma$ to obtain the densities $n_\sigma = n/2$, even in the
unpolarized case; see below. It is convenient to use 
$\mu =  (\mu_\dn + \mu_\up)/2 \ \ {\rm and} \ \  h =( \mu_\dn - \mu_\up )/2$
in place of the $\mu_\sigma$, and also to define the reduced mass $m/2$ and $0\le \gamma < 1$ given by 
\begin{equation}
{m \over 2} = {m_\up m_\dn \over{m_\up + m_\dn}} \ \ {\rm and} \ \ \gamma = \frac{m_\up - m_\dn}{m_{\up} + m_{\dn}}.
\label{m-and-gamma}
\end{equation}

We briefly describe our formalism~\cite{Paper1}
to introduce notation and highlight the changes arising from $m_\up \ne m_\dn$.
The partition function $Z(T,\mu_\sigma)$ is written as a functional integral over Grassmann fields.
Introducing a Hubbard-Stratonovich field $\Delta(x)$ which couples to $\sib_{\up}(x)\sib_{\dn}(x)$ and 
integrating out the fermions we obtain
$Z = \int D\Delta D\Ds e^{-S}$ with the action
$S = \int dx \left(|\Delta(x)|^2/g - {\rm Tr}\ln \gr^{-1}[\Delta(x)]\right)$.
The integration is over $x = ({\bf x}, \tau)$ where the imaginary time $0 \le \tau \le \beta = 1/T$.
The inverse Green's function is 
$\gr^{-1} = \left[(-\partial_{\tau} - h - \gamma\nabla^2/2m )\hat{1}
+ (\nabla^2/2m  +\mu )\hat{\tau}_3 + \Delta\hat{\tau}^{+} + \Ds \hat{\tau}^{-} \right]$
$\times \delta(x-x')$, and the
trace acts on Nambu-Gorkov space. 

Mean field (MF) theory corresponds to a uniform, static saddle point $\Delta_0$ for $Z$.
We find the gap equation $\delta S_0 / \delta\Delta_0 = 0$ with
$S_0 = S[\Delta_0]= {\beta\Delta_0^2/g} - \sum_{\kk,ik_n} {\rm Tr} \ln \gr_0^{-1}(k)$.
Here
$\gr_0^{-1}(k) = (ik_n-h +\gamma\epsilon)\hat{1} - \xi\hat{\tau}_3 + \Delta_0\hat{\tau}_1$
with fermionic Matsubara frequencies $ik_n = i(2n+1)\pi/\beta$.
To simplify notation we omit $\kk$ labels on all energies and write them as 
$\epsilon = |\kk|^2/2m$ and $\xi = \epsilon-\mu$.

The fermionic excitation energies, determined from the poles of $\gr_0(\kk,\omega+i0^+)$,
are
\begin{equation}
E_{1, 2} =  E \mp (\gamma\epsilon - h)
\ \ \ {\rm with} \ \ \  E = (\xi^2 + \Delta_0^2)^{1/2}. 
\eeq
The thermodynamic potential $\Omega = - \ln Z/\beta$ in MF theory is
given by $\Omega_0 = S_0/\beta  = \Delta_0^2/g + \sum_{\bf k} 
[\xi - E + E_{1}\Theta(-E_{1})+ E_{2}\Theta(-E_{2})]$ at $T=0$.

Precisely \emph{at} $T=0$, given $\mu, \Delta_0$ and $\gamma$, 
there is a range of $h$ values for which we obtain 
an unpolarized solution $n_\up = n_\dn$ with $E_{1, 2} > 0$.
This range is ${\rm max}( \gamma\epsilon - E) = A^{(-)}
\le h \le A^{(+)}= {\rm min}(\gamma\epsilon + E)$, where the max/min are for $\epsilon \ge 0$.  
It is easy to see that $A^{(\pm)} = \gamma \mu \pm \sqrt{1-\gamma^2} \Delta_0$ for $\mu/\Delta_0 \ge \pm \Gamma$ 
while $A^{(\pm)} = \pm\sqrt{\mu^2 + \Delta_0^2}$ for $\mu/\Delta_0 < \pm \Gamma$, where 
$\Gamma  \equiv  \gamma/\sqrt{1-\gamma^2}$. In the limit $T \rightarrow 0$
there is a unique solution~\cite{footnote} $h = [A^{(+)}+A^{(-)}]/2$;
deviations away from this value lead to an exponentially small polarization at low $T$ within the interval. 

The expression for $S_0$ simplifies greatly with $E_{1, 2} > 0$ and we
obtain the gap equation 
\beq 
-m/ 2 \pi a_s = \sum_{\kk} \left(E^{-1 } - \epsilon^{-1} \right) 
\label{gap_equation}
\eeq
and MF number equation
$n = -\partial\Omega_0/\partial\mu = \sum_{\kk} \left(1 - {\xi / E} \right)$.
Thus the properties of the $m_\up \ne m_\dn$ system at the MF level
are completely equivalent to those of an equal mass system with 
chemical potential $\mu =  (\mu_\dn + \mu_\up)/2$ and $m$ twice the reduced mass (\ref{m-and-gamma}). 
This was already noted in ref.~\cite{SadeMelo}; here we go beyond their MF analysis 
by including the effect of Gaussian fluctuations on the equation of state.

We write $\Delta(x)  = \Delta_0 + \eta (x)$ where the $\eta$ are $({\bf x},\tau)$ dependent fluctuations
about the saddle point $\Delta_0$. Fourier transforming to momentum $\bf q$ and bosonic frequencies $iq_\ell = i2\pi \ell/\beta$ and 
expanding to quadratic order in $\eta$ we get 
$S \simeq S_0 + S_g$, where
$S_g = {1 \over 2}\sum_{\qq,iq_l} \eta^\dag \MM \eta$ with $\eta^\dag = \left( \eta^*(q),\eta(-q) \right)$
The inverse fluctuation propagator $\MM$  is given by 
$\MM_{11}(q)  = \MM_{22}(-q) = 1/g + \beta^{-1}\sum_{\kk,ik_n} \gr^0_{22}(k)\gr^0_{11}(k+q)$ and
$\MM_{12}(q) = \MM_{21}(q) = \beta^{-1}\sum_{\kk,ik_n} \gr^0_{12}(k)\gr^0_{12}(k+q)$. At $T=0$ we get
\bqa
\MM_{11}(q)&=&\frac{1}{g} + \sum_\kk \left[\frac{u^2u'^2}{iq_\ell-E_1-E_2'} - \frac{v^2v'^2}{iq_\ell+E_1+E_2'}\right]  \no\\
\MM_{12}(q)&=&\sum_\kk \left[ \frac{uvu'v'}{iq_\ell+E_1+E_2'} 
 -\frac{uvu'v'}{iq_\ell-E_1-E_2'}\right].
\label{m12}
\eqa
Here we use standard BCS notation 
$v^2 = 1 - u^2 = {1 \over 2}\left( 1 - \xi / E \right)$ 
together with the abbreviations $u = u_\kk, u' = u_{\kk + \qq}$, etc.
The only difference in $\MM_{ij}$ from the equal mass case is the presence
of $E_1+E_2' = E+E'-\gamma(\epsilon' - \epsilon)$. 
Thus, even at the Gaussian level, $h$ does \emph{not} enter the calculation so long as the system is unpolarized. 

\begin{figure}[t!]
\includegraphics[width=3in]{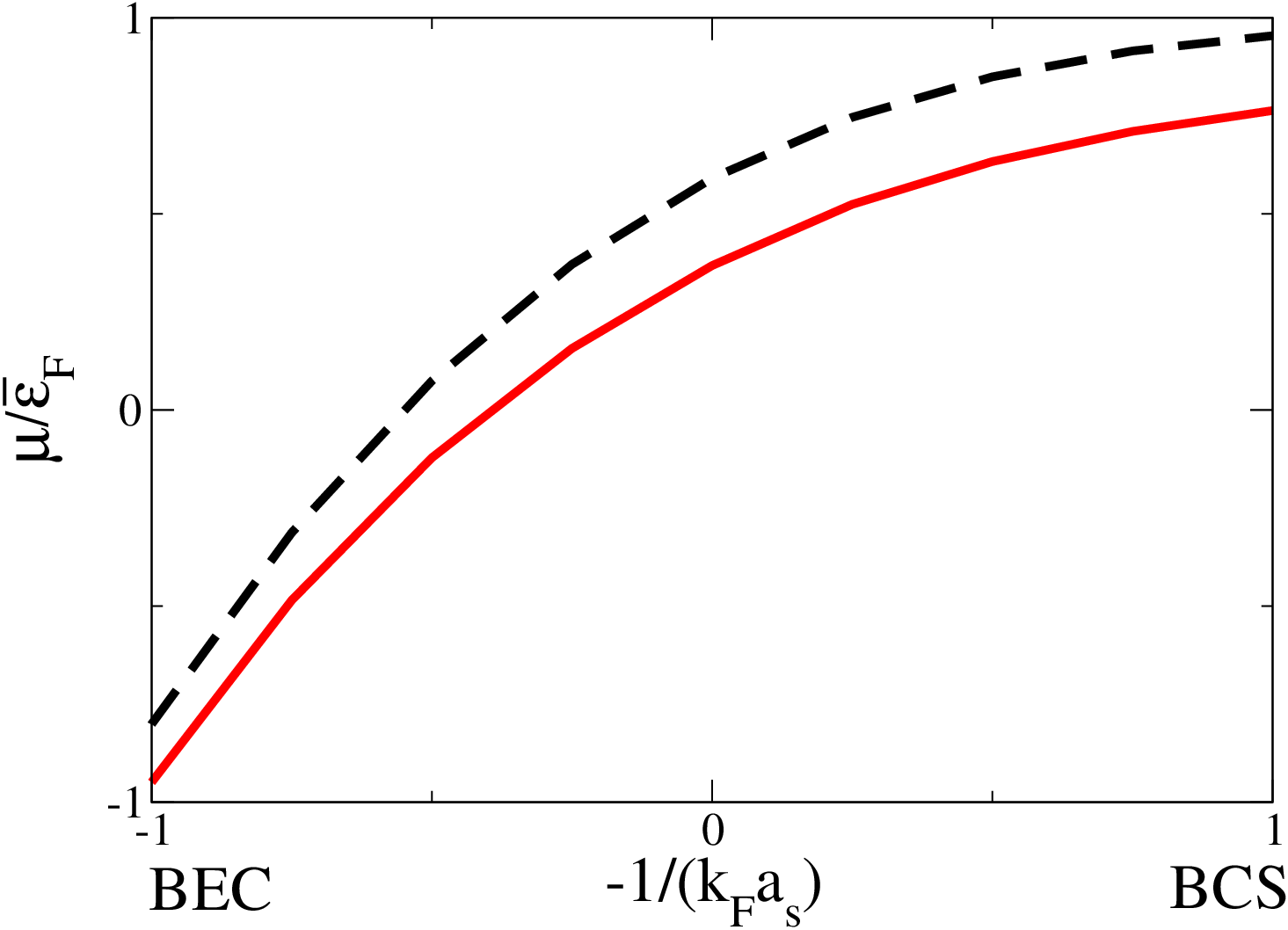}
\caption{Chemical potential $\mu$ normalized by
$\overline{\epsilon}_F = (\epsilon_{F\up} +\epsilon_{F\dn})/2$ as a function of the interaction $-1/(k_Fa_s)$ with $m_\up/m_\dn = 6.67$ for $^{40}$K and $^6$Li.
The solid (red) line includes Gaussian fluctuations, while the dashed (black) line is the mean field prediction (independent of $m_\up/m_\dn$).}
\end{figure}

The partition function is now approximated by
$Z \simeq \exp({-S_0}) \, \int D\eta D\eta^\dagger \, \exp(-S_g)$.
As explained in ref.~\cite{Paper1,Polar paper} the saddle point equation still retains its earlier form
$\delta S_0/\delta \Delta_0 = 0$, but the thermodynamic potential $\Omega$ at $T=0$ now includes the contributions of the energy of zero point motion of the collective excitations (the Goldstone mode of the broken $U(1)$ symmetry in the superfluid) and the virtual scattering of gapped fermionic quasiparticles.
A careful consideration of convergence factors~\cite{Paper1} leads to $\Omega \simeq  \Omega_0 + \Omega_g + \ldots$ with
\beq
\Omega_g =  
\frac{1}{2\beta} \sum_{\qq,iq_\ell} \ln \left[\MM_{11}(q){\rm Det}\MM (q)/\MM_{22}(q) \right]e^{iq_\ell 0^+}.
\label{omega_g}
\eeq
The Gaussian contribution leads to a lowering of the ground state energy and a modified equation of state which we calculate as $n = -{\partial  \Omega / \partial \mu}$
at $T=0$. Given the dependence of $\Delta_0$ on $\mu$, we thus find 
\beq
n=-\partial \Omega_0/\partial\mu - \partial\Omega_g[\mu,\dl(\mu)]/\partial\mu.
\eeq
In Fig.~1 we show the average chemical potential $\mu$ as a function of the scattering length $a_s$
for the experimentally relevant case of a $^{40}$K-$^6$Li mixture. 

To understand in greater detail how unequal masses alter the results from the well studied equal mass case, we next concentrate on two regions: unitarity and the BEC limit.
It is also useful to compare our results with those of a closely related approach,
where one generalizes the Hamiltonian (\ref{hamiltonian}) to ${\cal N}$ flavors of fermions of each species,
interacting with an Sp($2{\cal N}$)-symmetric potential~\cite{1overN}.  
Then MF theory becomes exact in the ${\cal N}\rightarrow \infty$ limit 
and the Gaussian correction is of order $1/{\cal N}$ so that 
$\Omega/{\cal N} = \Omega_0 +  \Omega_g/{\cal N} + {\cal O} (1/{\cal N}^2)$.
Thus, we write $\mu = \mu_0 + \mu_1/{\cal N}$ where $\mu_0$ is the MF value obtained from
$n = -\partial \Omega_0 / \partial \mu$, and 
$\mu_1 = -(\partial \Omega_g / \partial \mu)/(\partial^2 \Omega_0 / \partial \mu^2)$ evaluated at $\mu_0$~\cite{Comment on 1overN}.
In order to obtain results relevant to the original Hamiltonian ${\cal N} = 1$ is set at the end of the calculation. 
In contrast, in our Gaussian theory we work directly with the Hamiltonian (\ref{hamiltonian}),
keeping ${\cal N}=1$ from the outset, and calculate corrections to various quantities 
(like the chemical potential) without assuming that the changes are small.

\begin{figure}[t!]
\includegraphics[width=3in]{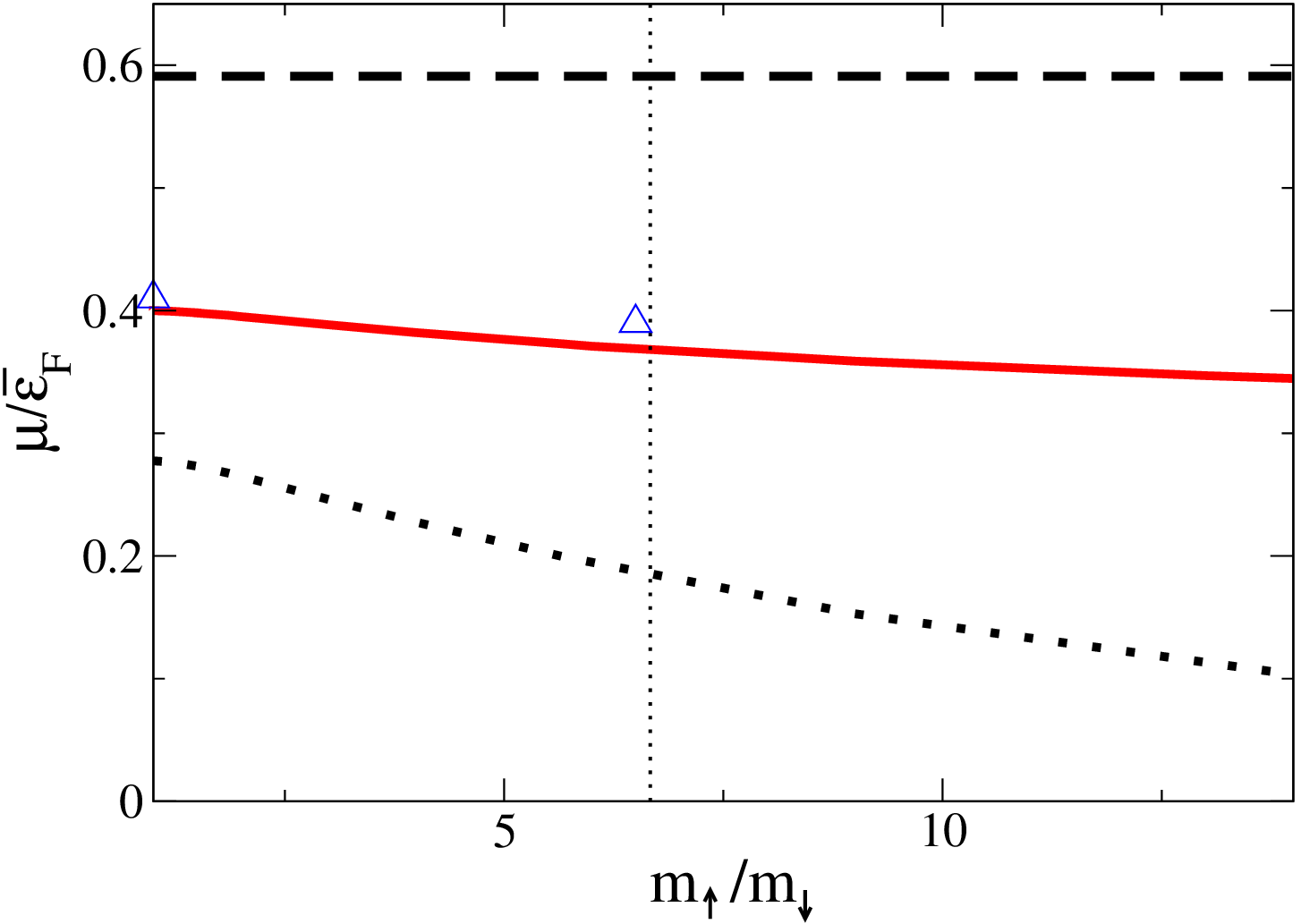}
\caption{The chemical potential $\mu$ at unitarity as a function of the mass ratio.
At unitarity, the ground state energy $E$ and $\mu$ are related by universality:
$\mu/\overline{\epsilon}_F = E/(3N\overline{\epsilon}_F/5) = (1+\beta)$
where $\overline{\epsilon}_F = (\epsilon_{F\up} +\epsilon_{F\dn})/2$.
Our calculation including Gaussian fluctuations is the solid red line. The
vertical line marks $m_\up/m_\dn = 6.67$ for a $^{40}$K/$^6$Li mixture. 
For comparison, the two triangles are Quantum Monte Carlo results~\cite{Carlson},  
the upper dashed line is the mean field result and the lower
dotted line that of the $1/{\cal N}$ expansion.}
\end{figure}

At unitarity $a_s= \infty$ and the interactions have no scale; the chemical potential is the only energy available~\cite{universal}.  
The saddle point condition implies $\mu= 0.860\,\Delta_0 $ independent of mass ratio and  
the thermodynamic potential has the universal scaling form
\beq
\Omega(\mu) = {- 2\over 15 \pi^2} \, {\cal F}(m_\up/m_\dn)\,  \mu^{5/2} (2m)^{3/2} 
\eeq
where the function ${\cal F}$ is normalized so that ${\cal F}(1) = 1$
for \emph{non-interacting} fermions with \emph{equal} masses. 

In the MF approximation ${\cal F}_0 = 2.2032$, independent of $m_\up/m_\dn$, 
which yields the equation of state $\mu_0 = \overline{\epsilon}_F/{\cal F}_0^{2/3}$ where 
$\overline{\epsilon}_F = k_F^2/(2m) = (\epsilon_{F\up} + \epsilon_{F\dn})/2$.  
The Gaussian contribution ${\cal F}_g$ has a non-trivial, though weak, dependence on the mass ratio 
obtained by evaluating (\ref{omega_g}) numerically.
This yields $\mu = \overline{\epsilon}_F/[{\cal F}_0 + {\cal F}_g(m_\up/m_\dn)]^{2/3}$.
This result is plotted in solid line in Fig.~2, with the dashed line showing the mean-field result. 
The dotted line corresponds to the $1/{\cal N}$ result 
$\mu = [1- 2{\cal F}_g(m_\up/m_\dn)/3{\cal N}{\cal F}_0]\mu_0$ with ${\cal N}=1$.  
In Fig.~2 we also show the result of a recent Quantum Monte Carlo calculation~\cite{Carlson} 
(triangles) which agrees fairly well with our approximation.

\begin{figure}[t!]
\includegraphics[width=3in]{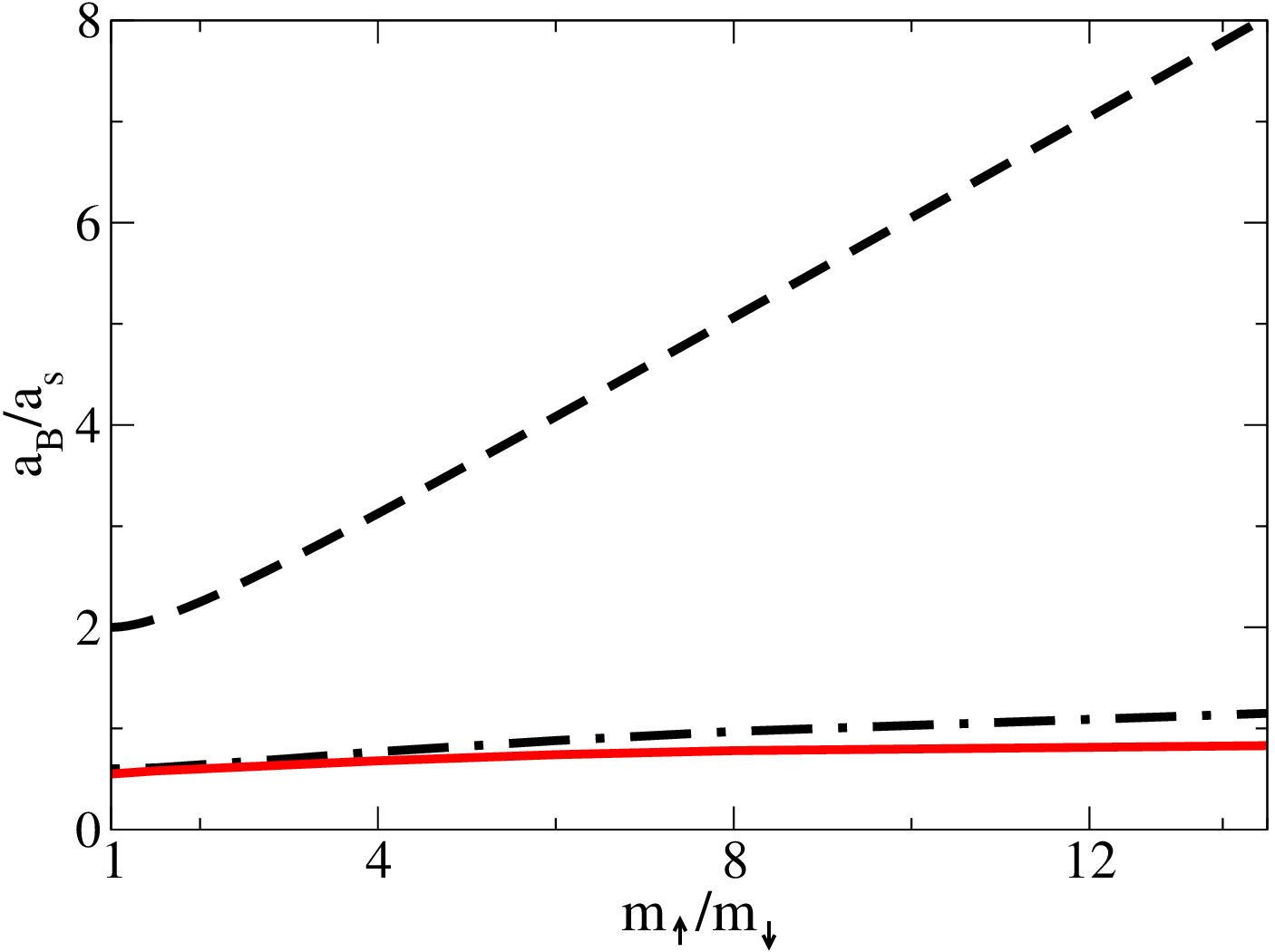}
\caption{Molecular scattering length in the BEC limit as a function of mass ratio for different calculations.  Our calculation including Gaussian fluctuations is the solid red line; the exact 4-body calculation~\cite{Petrov} is shown as the dot-dash line. The mean field value, corresponding to the Born approximation for dimer scattering, is the upper dashed line.}
\end{figure}

Let us next discuss the BEC limit ($a_s \rightarrow 0^+$) where we obtain a condensate of tightly bound diatomic molecules of 
binding energy $E_B = 1/(ma_s^2)$, mass $m_B = m_\up + m_\dn$ and density $n_B = n/2$. These bosonic dimers
are weakly interacting with a (dimer) scattering length $a_B$, which we now compute.
We find it convenient to define dimensionless variables $x$ and $y$
as $|\mu| = (1 - y)/(2ma_s^2)$, with $0 \le y \ll 1$ and
$(2ma_s^2\Delta_0)^2 = x \ll 1$ in the BEC limit. We can thus expand to quadratic order  
${S_0(x,y)/\beta} \simeq [-xy + x^2/16]/(32\pi m a_s^5)$. The gap equation
$\partial S_0/\partial x = 0$ then yields $x = 8y$. 

At the MF level $\Omega_0(\mu) ={S_0(x=8y,y)/\beta}$ so that
the number equation $n = - (2ma_s^2)\partial\Omega_0/\partial y$ yields $y = 2\pi n a_s^3$.
We find $\mu \simeq  -1/(2ma_s^2) + \pi n a_s/m$ and on identifying the second term
with $\mu_B/2 = 2\pi n_B a_B/m_B$ we obtain the MF result~\cite{SadeMelo} 
$a_B^{\rm MF}/a_s = (m_\up + m_\dn)/m$. This
corresponds to the Born approximation for dimer scattering and
is shown as a dashed line in Fig.~3.

The Gaussian contribution (\ref{omega_g}) in the BEC limit can be written as
$\Omega_g = - x^2 g(m_\up/m_\dn) / [512 \pi m a_s^5]$
where the function $g(m_\up/m_\dn)$ can be numerically calculated by 
the method of ref.~\cite{Paper1} (Appendix E).
Using $x = 8y$, we write $\Omega(\mu) = \Omega_0 + \Omega_g = - (1 + g)y^2/(8\pi m a_s^5)$.
Solving the number equation $n = - (2ma_s^2)\partial\Omega/\partial y$
for $\mu$ we determine the dimer scattering length (as explained above) to
find~\cite{1/Nab} $a_B/a_s = (m_\up + m_\dn)/[(1 + g(m_\up/m_\dn))m]$.
This is plotted as the solid line in Fig.~3 and compares very well with the exact result 
from a 4-body calculation of the scattering between dimers~\cite{Petrov}. 

We use our numerical results to
find the next correction in the BEC limit, by fitting $\Omega(\mu)$ at small $y$ to 
to the form $(Ay^2 + By^{5/2} + \ldots)/(m a_s^5)$.
We thus find the $T=0$ equation of state of the Lee-Yang-Huang form
\beq
\mu_B = {4\pi n_B a_B \over m_B} [1 + 32C(n_B a_B)^{3/2}/(3\sqrt{\pi}) + \dots]
\eeq
arising from the quantum depletion of the condensate.
From our numerics we find $C = 0.90 \pm 0.05$ over the mass ratio range 
$1 \le m_\up/m_\dn \le 13.6$. 
The exact result~\cite{Lee-Yang-Huang} for the dilute Bose gas is $C = 1$.

Finally we analyze the effect of a harmonic trap within the Thomas-Fermi local density approximation (LDA); for simplicity we discuss spherical traps. 
The two species will in general see \emph{different} trapping potentials with
frequencies $\omega_\sigma$, so that within LDA 
$\mu_\sigma(r) = \overline{\mu_\sigma} - m_\sigma\omega^2_\sigma r^2/2$. 
Let us focus on unitarity where $\Delta_0(r)/\mu(r) = 1.162 \equiv \delta$ for all $r$.
To ensure zero polarization at $T=0$ we must satisfy $h(r)/\mu(r) = K$. 
Using the results derived below eq.~(3), $K =\gamma$ for $\gamma/\sqrt{1-\gamma^2} < 1/\delta$ corresponding
to $m_\up/m_\dn < 4.74$, while 
$K = [\gamma - \delta\sqrt{1-\gamma^2} + \sqrt{1+\delta^2}]/2$ for  $m_\up/m_\dn > 4.74$.  
To maintain this ratio of $h/\mu$ we must choose
\begin{equation}
{\omega_\uparrow^2 \over \omega_\downarrow^2} = {m_\downarrow (1-K)\over m_\uparrow (1+K)} 
\end{equation}
Thus, $\omega_\uparrow/ \omega_\downarrow = m_\downarrow / m_\uparrow$ for  $m_\up/m_\dn < 4.74$ while for a Li-K mixture, where $K = 0.745$, we get 
$\omega_\up/\omega_\dn  \simeq 0.148$.
Deviations from this frequency ratio would lead to regions of spin imbalance in the trap.


We conclude with a brief discussion of some general issues. First, note that mass imbalance does not impact 
pairing of ``time-reversed'' states. For the unpolarized case the Fermi spheres match in $\kk$-space
even if the $\mu$'s are different. 
Unlike in many condensed matter systems where time-reversal breaking perturbations~\cite{Maki} 
flip the spin of the fermions, atomic species retain their identity (``spin'') in collisions. 
However, effects beyond pairing (which is what we focused on) may well destabilize the system.
In the BEC limit, it is known that for $m_\up/m_\dn > 13.6$ the presence of Efimov states
makes the Feshbach molecules unstable to collapse by three-body recombination~\cite{Petrov}. 
Finally, we may ask why our approach seems to work so well 
when compared with available exact results (at least in the stable mass ratio regime). 
The answer seems to lie in the fact that in a homogeneous dilute gas (with $k_F^{-1} \gg$ range of potential) 
particle-hole channel contributions are negligible compared to 
the particle-particle channel ones that our approach is designed to capture. 

We gratefully acknowledge support from ARO W911NF-08-1-0338 and NSF-DMR 0706203.


\begin{thebibliography}{99}
\bibitem{footnote}
The MF result $\sum_\kk[\tanh(\beta E_1/2) - \tanh(\beta E_2/2)] = 0$
can be rewritten as $\exp(2\beta h) = I^{(-)}/I^{(+)}$, where 
$ I^{(\pm)} = \int_0^\infty d\epsilon \sqrt{\epsilon} \, \exp[-\beta(E \pm \gamma\epsilon)]$.
We find $h = [A^{(+)}+A^{(-)}]/2$ by evaluating $ I^{(\pm)} \approx \exp[\mp \beta A^{(\pm)}]$
for $T\rightarrow 0$, since the integral is dominated by the region where the exponent reaches its maximum. 
At the Gaussian level $\MM$ is independent of $h$; thus $\partial\Omega/\partial h = \partial\Omega_0/\partial h$
and the MF result for $h$ remains valid.
\bibitem{Comment on 1overN}
This equation of state is equivalent~\cite{Polar paper} to the result of ref.~\cite{1overN}.
\bibitem{1/Nab}
We show elsewhere~\cite{Polar paper} that the $1/{\cal N}$ result for the dimer scattering length
at ${\cal N}=1$ is the same as the Gaussian result reported here.
\end{thebibliography}
\end{document}